\documentclass{SCGE}
\usepackage[dvipdfmx,bookmarks=true,colorlinks,%
citecolor=blue,linkcolor=blue,anchorcolor=blue,filecolor=blue,urlcolor=blue]{
hyperref}

\begin{document}
\bibliographystyle{sci-chin}

\begin{picture}(0,0){\rm
\put(0,-20){\makebox[160truemm][l]{\bf {\sanhao\raisebox{2pt}{.}}
Article  {\sanhao\raisebox{1.5pt}{.}}}}}
\put(0,-34){\jiuwuhao {\textcolor[rgb]{0.5,0.5,0.5}{\sf 
}}}
\end{picture}

\def\bm{\boldsymbol}

\def\dl{\displaystyle}
\def\du{\end{document}}
\def\d{{\rm d}}
\def\e{{\rm e}}
\def\i{{\rm i}}
\def\pi{{\uppi}}

\Year{2015} %
\Month{??} %
\Vol{58} 
\No{?} 
\BeginPage{1} 
\AuthorMark{{\rm Wang B}, et al.}  
\AuthorMarkCite{{\rm Wang B, Zhao W J, Zhao E G, Zhou S G}. } 
\DOI{??} 

\title[Theoretical study of fusion reactions 
$^{32}$S + $^{94,96}$Zr and 
$^{40}$Ca + $^{94,96}$Zr 
and quadrupole deformation of $^{94}$Zr]
{Theoretical study of fusion reactions 
$^{32}$S + $^{94,96}$Zr and 
$^{40}$Ca + $^{94,96}$Zr 
and quadrupole deformation of $^{94}$Zr}

\author[1]{WANG Bing}{}
\author[1]{ZHAO WeiJuan}{}
\author[2,3]{ZHAO EnGuang}{}
\author[2,3,4*]{ZHOU ShanGui}{}

\address[{\rm1}]{Department of Physics, Zhengzhou University, Zhengzhou 450001, 
                 China;}
\address[{\rm2}]{State Key Laboratory of Theoretical Physics, Institute of 
                 Theoretical Physics, Chinese Academy of Sciences, Beijing 
                 100190, China;}
\address[{\rm3}]{Center of Theoretical Nuclear Physics, National Laboratory
                 of Heavy Ion Accelerator, Lanzhou 730000, China}
\address[{\rm4}]{Synergetic Innovation Center for Quantum Effects 
and Application,
              Hunan Normal University, Changsha, 410081, China}

\maketitle \vspace{-3.5mm}{\footnotesize\begin{center} Received XX, 2015; 
accepted XX, 2015; published online XX, 2015
\end{center}}\vspace*{-5mm}

\begin{center}
\rule{16.5cm}{0.4pt}
\parbox{16.5cm}
{\begin{abstract} 
The dynamic coupling effects on fusion cross sections for reactions 
$^{32}$S + $^{94,96}$Zr and $^{40}$Ca + $^{94,96}$Zr are studied with the 
universal fusion function formalism and an empirical coupled channel (ECC) model. 
An examination of the reduced fusion functions shows that the total effect of 
couplings to inelastic excitations and neutron transfer channels on 
fusion in $^{32}$S + $^{94}$Zr ($^{40}$Ca + $^{94}$Zr) is almost the same 
as that in $^{32}$S + $^{96}$Zr ($^{40}$Ca + $^{96}$Zr). 
The enhancements of the fusion cross section at sub-barrier energies due 
to inelastic channel coupling and neutron transfer 
channel coupling are evaluated separately by using the ECC model. 
The results show that effect of couplings to inelastic excitations channels 
in the reactions with $^{94}$Zr as target should be similar as that 
in the reactions with $^{96}$Zr as target. 
This implies that the quadrupole deformation parameters $\beta_2$ of $^{94}$Zr and 
$^{96}$Zr should be similar to each other. 
However, $\beta_2$'s predicted from 
the finite-range droplet model, which are used in the ECC model, are quite different. 
Experiments on $^{48}$Ca + $^{94}$Zr or $^{36}$S + $^{94}$Zr are suggested 
to solve the puzzling issue concerning $\beta_2$ for $^{94}$Zr.
\end{abstract}}
\end{center}\vspace*{-0.6cm}

\begin{center}
\parbox{16.5cm}
{\bf\jiuhao Empirical coupled channel model, barrier distribution, 
universal fusion function, neutron transfer}
\end{center}

\begin{center}
{}

\CITA    
\end{center}

\textwidth=178truemm \textheight=236truemm

\wuhao\vspace*{1.5mm}

\begin{multicols}{2}

\renewcommand{\baselinestretch}{1.08} \baselineskip 12.2pt\parindent=10.8pt

\renewcommand{\thefootnote}

\section{Introduction}\label{sec:intro}

\label{sec:introduction}

Heavy-ion fusion reaction has been an interesting 
topic for several decades because the heavy-ion fusion not only is of central 
importance for nucleosynthesis but also can reveal rich interplay between 
nuclear structure and reaction dynamics 
\cite{Balantekin1998_RMP70-77,Dasgupta1998_ARNPS48-401,Canto2006_PR424-1,
Back2014_RMP86-317,Meissner2014_SciBulletin60-43,Canto2015_PR596-1,
Fu2015_SciBulletin60-1211}.  
The study of fusion reaction mechanism is also of fundamental importance 
for understanding the synthesis of superheavy elements,
properties of weakly bound nuclei,
and symmetry energy of the nuclear equation of state
\cite{Adamian1997_NPA618-176,Adamian1998_NPA633-409,Adamian1999_PLB451-289,
Knyazheva2007_PRC75-064602,Wang2012_PRC85-041601R,Wang2014_JPCS515-012022,
Zhang2013_NPA909-36,
Shen2014_SciChinaPMA57-453,
Fan2015_SciChinaPMA58-062002,
Mo2015_SciChinaPMA58-082001}
Up to now, lots of important information about fusion dynamics at energies 
near the Coulomb barrier, 
especially at sub-barrier energies, are obtained through experimental and 
theoretical studies, such as the fusion hindrance phenomenon at extreme low 
energies---a steep falloff of the fusion cross sections 
\cite{Jiang2002_PRL89-052701,Misicu2006_PRL96-112701,Dasgupta2007_PRL99-192701,
Diaz-Torres2008_PRC78-064604,Ichikawa2009_PRL103-202701,
Denisov2014_PRC89-044604}, the role of the neutron transfer effect in the 
%
fusion \cite{Gomes2011_PRC84-014615,Sargsyan2012_PRC86-054610,
Gasques2009_PRC79-034605,Dasgupta2010_NPA834-147c}, the breakup effect on the 
fusion reactions process 
\cite{Beckerman1980_PRL45-1472,Broglia1983_PRC27-2433R,Broglia1983_PLB133-34,
Stelson1990_PRC41-1584,Dasso1983_NPA405-381,Dasso1983_NPA407-221,
Zhang2014_PRC89-054602,Wang2014_PRC90-034612}, etc..
 
In the sub-barrier energy region, a large enhancement of fusion cross section 
for the fusion reaction of $^{58}$Ni + $^{64}$Ni as compared with that for 
$^{58}$Ni + $^{58}$Ni and $^{64}$Ni + $^{64}$Ni was observed by Beckerman {\it 
et al.} \cite{Beckerman1980_PRL45-1472}. Broglia {\it et al.} 
\cite{Broglia1983_PRC27-2433R,Broglia1983_PLB133-34} suggested that the 
coupling to transfer channels with positive $Q$ values is needed to explain the 
enhancement of fusion data for the Ni + Ni systems. Large enhancements of 
sub-barrier fusion cross sections have been also observed in 
many other reaction systems with positive $Q$-value neutron transfer (PQNT) 
channels, such as the reaction systems $^{32}$S + $^{A}$Pd ($A$ = 104--106, 
108, \text{and} 110) \cite{Pengo1983_NPA411-255}, $^{40}$Ca + $^{44,48}$Ca 
\cite{Aljuwair1984_PRC30-1223}, $^{40}$Ca + $^{94,96}$Zr 
\cite{Timmers1998_NPA633-421,Stefanini2007_PRC76-014610}, and $^{32}$S
+ $^{94,96}$Zr \cite{Zhang2010_PRC82-054609,Jia2014_PRC89-064605}. For some of 
these systems, the fusion excitation functions have been measured in 
sufficiently small energy steps, which can be used to extract the underlying 
barrier distributions to study the contribution from transfer channels. The 
experimental barrier distributions are much broader than those of the 
reaction systems with negative $Q$-value neutron transfer channel. However, in 
some other experiments for reaction systems 
with PQNT channels \cite{Jacobs1986_PLB175-271,Jia2012_PRC86-044621}, no 
enhancement was observed in the 
fusion cross sections at sub-barrier energies. 
 
Theoretically, many efforts have been made to understand the effect of the 
neutron transfer channels. In Ref.~\cite{Stelson1990_PRC41-1584}, the authors 
proposed that a neutron flow between the projectile and the target nuclei 
before fusion could promote neck formation which provides a force strong enough 
to overcome the Coulomb force. Therefore the fusion is more favored and the 
sub-barrier fusion cross section is enhanced. In 
Ref.~\cite{Zagrebaev2003_PRC67-061601R}, 
Zagrebaev proposed a simplified semiclassical model to describe the effect of 
neutron transfer on fusion. This effect depends on both neutron transfer 
probabilities and their $Q$ values. The PQNT provides a gain in the 
kinetic energy. Consequently, the fusion is easier and the 
sub-barrier fusion cross section is enhanced. Sargsyan {\it et al.} suggested 
that the deformations of the interacting nuclei change owing to the PQNT
\cite{Sargsyan2012_PRC86-014602,Sargsyan2013_PRC88-064601,
Sargsyan2015_PRC91-014613}. Thus, the influence of the PQNT channels on fusion 
is accompanied by and depends on the change of nuclear deformations.
In the quantum coupled channel (QCC) model \cite{Hagino1999_CPC123-143}, the 
coupling to PQNT channels is treated approximately by using a macroscopic
form factor. Within the microscopic dynamics models, such as the quantum 
molecular dynamic model 
\cite{Wang2002_PRC65-064608,Wang2014_PRC90-054610,Wang2004_PRC69-034608,
Wen2013_PRL111-012501,Wen2014_PRC90-054613,Wang2015_SciChinaPMA58-112001} and the 
time-dependent Hartree-Fock
method 
\cite{Umar2006_PRC74-021601R,Umar2012_PRC85-017602,Keser2012_PRC85-044606, 
Guo2007_PRC76-014601,Guo2008_PRC77-041301R, 
Dai2014_SciChinaPMA57-1618,Dai2014_PRC90-044609}, 
the effects of surface excitations as well as nucleon transfer can be 
automatically included. Up to now, although many experiments and theoretical 
efforts have been devoted to study the mechanism of the coupling to PQNT 
channels, the underlying mechanism is still far from a clear understanding.

In our previous work 
\cite{Wang2015_arXiv1504.00756}, a systematic study of 
capture excitation functions for 217 reaction systems has been performed by 
using an empirical coupled channel (ECC) model. In the ECC model, a barrier 
distribution is used to take effectively into account the effects of 
couplings. The effect of the coupling to PQNT channels is simulated by 
broadening the barrier distribution. Among these 217 reaction systems, there 
are 86 systems with positive $Q$ values for one neutron pair transfer 
channel. The calculated capture cross sections of most of these 86 reaction 
systems are in good agreement with the experimental values, including 
$^{32}$S + $^{96}$Zr and $^{40}$Ca + $^{96}$Zr. However, the calculated results 
underestimate the sub-barrier cross sections for $^{32}$S + $^{94}$Zr and 
$^{40}$Ca + $^{94}$Zr. These results seem to be similar to those obtained 
from Ref.~\cite{Jia2014_PRC89-064605}. In Ref.~\cite{Jia2014_PRC89-064605}, a 
large enhancement for the sub-barrier fusion cross sections was deduced in 
$^{32}$S + $^{94}$Zr compared to $^{32}$S + $^{96}$Zr based on QCC 
calculations without the neutron transfer effect considered, although the 
$Q(xn)$ values, which are listed in Table \ref{tab:Qval}, for $^{32}$S + 
$^{94}$Zr are relatively smaller than those for $^{32}$S + $^{96}$Zr. The 
authors suggested that the 
neutron transfer effect in $^{32}$S + $^{94}$Zr are much stronger than that in 
$^{32}$S + $^{96}$Zr. In the present work, we will study the dynamic 
coupling effects in the reactions $^{32}$S + $^{94,96}$Zr and $^{40}$Ca + 
$^{94,96}$Zr with the universal fusion function (UFF) formalism and the ECC 
model. We will first investigate the dynamic coupling effects on fusion 
cross sections with the UFF formalism. Then, the 
effects of couplings to inelastic excitations and neutron transfer channels on 
fusion are analysed separately with the ECC model.

The present paper is organized as follows. In Sec.~\ref{sec:methods}, the ECC 
model is briefly reviewed. The method used to eliminate geometrical factors 
and static effects of the data is introduced in Sec.~\ref{sec:results} where the 
influence on fusion cross section owing to inelastic excitations and transfer 
couplings will be also investigated. A summary is given in 
Sec.~\ref{sec:summary}.

\vspace*{-5mm} 
\begin{table}[H]
\caption{$Q$ values for one or multineutron transfer channels
from ground state to ground state for $^{32}$S + $^{90,94,96}$Zr, 
$^{36}$S + $^{94}$Zr, $^{40}$Ca + $^{90,94,96}$Zr, and  $^{48}$Ca + 
$^{90,94,96}$Zr.}
\label{tab:Qval}
\begin{center}\footnotesize \doublerulesep 0.2pt
\begin{tabular*}{0.38\paperwidth}{@{\extracolsep{\fill}}lrrrr}
\toprule
  Reaction  & $Q(1n)$ & $Q(2n)$ & $Q(3n)$ & $Q(4n)$ \\
   & (MeV) & (MeV) & (MeV) & (MeV) \\\hline 
 $^{32}$S$+^{90}$Zr  & $-$3.33 & $-$1.23 & $-$6.60 & $-$6.16   \\
 $^{32}$S$+^{94}$Zr  &    0.42 &    5.10 &    3.46 &    6.15   \\
 $^{32}$S$+^{96}$Zr  &    0.79 &    5.74 &    4.51 &    7.66   \\
 $^{36}$S$+^{94}$Zr  & $-$3.92 & $-$2.61 & $-$6.88 & $-$6.32   \\
 $^{40}$Ca$+^{90}$Zr & $-$3.61 & $-$1.44 & $-$5.86 & $-$4.18    \\
 $^{40}$Ca$+^{94}$Zr &    0.14 &    4.89 &    4.19 &    8.12   \\
 $^{40}$Ca$+^{96}$Zr &    0.51 &    5.53 &    5.24 &    9.64    \\
 $^{48}$Ca$+^{90}$Zr & $-$6.82 & $-$9.78 & $-$17.31& $-$20.77    \\
 $^{48}$Ca$+^{94}$Zr & $-$3.07 & $-$3.45 & $-$7.26 & $-$18.53    \\
 $^{48}$Ca$+^{96}$Zr & $-$2.71 & $-$2.81 & $-$6.21 & $-$6.95    \\
\bottomrule
\end{tabular*}
\end{center}
\end{table} 
\vspace*{-5mm}

\section{Methods} \vspace*{-1mm}
\label{sec:methods}

The fusion cross section at a given center-of-mass energy $E_{\rm c.m.}$
can be written as the sum of the cross section for
each partial wave $J$, 
\begin{equation}\label{eq:sig_cap}
\sigma_{\rm f}(E_{\rm c.m.})=\frac{\pi\hbar^2}{2\mu E_{\rm c.m.}} 
                             \sum_{J}^{J_{\rm max}}(2J+1)T(E_{\rm 
c.m.},J), 
\end{equation}
here~$\mu$~denotes the reduced mass of the reaction system and $T$ denotes 
the penetration probability. $J_{\rm max}$
is the critical angular momentum: For the partial wave with angular
momentum larger than $J_{\rm max}$, the ``pocket'' of the interaction
potential disappears. The interaction potential around the Coulomb barrier is 
approximated by an ``inverted'' parabola. 

The couplings between the relative motion of 
the two nuclei and other degrees of freedom including the coupling to PQNT 
channels results in an enhancement in the fusion cross sections at sub-barrier 
energies. In the ECC model 
\cite{Wang2015_arXiv1504.00756}, a barrier distribution $f(B)$ is introduced 
to take into account the coupled channel effects in an empirical way. Then, the
penetration probability is calculated as 
\begin{equation}\label{eq:Tran}
  T(E_{\rm c.m.},J) =\int f(B)T_{\rm HW}(E_{\rm c.m.},J,B){\rm 
d}B. 
\end{equation}
$T_{\rm HW}$ denotes the penetration probability calculated by the well-known 
Hill-Wheeler formula \cite{Hill1953_PR089-1102}. 
Note that for very deep sub-barrier penetration, the Hill-Wheeler formula is not
valid because of the long tail in the Coulomb potential.
In Ref.~\cite{Li2010_IJMPE19-359}, a new barrier penetration formula was
proposed for potential barriers containing a long-range Coulomb interaction
and this formula is especially appropriate for the barrier penetration
with penetration energy much lower than the Coulomb barrier.

The barrier distribution $f(B)$ is taken to be an asymmetric
Gaussian function
 
\begin{equation}\label{eq:distri}
f(B)=\left\{
      \begin{array}{cc}
       \frac1N\exp\left[-\left(\frac{B-B_{\rm m}}{\varDelta_1}\right)^2\right],
                                                  \quad & B < B_{\rm m} \\[1em]
       \frac1N\exp\left[-\left(\frac{B-B_{\rm m}}{\varDelta_2}\right)^2\right],
                                                          \quad & B > B_{\rm m}
      \end{array}
     \right.. 
\end{equation}
$f(B)$ satisfies the normalization condition $\int
f(B)dB=1$. $N =\sqrt{\pi}(\varDelta_1+\varDelta_2)/2 $~is a normalization
coefficient. $\varDelta_1$, $\varDelta_2$, and $B_{\rm m}$ denote the left 
width, the right width, and the central value of the barrier distribution, 
respectively.

Within the ECC model \cite{Wang2015_arXiv1504.00756}, the barrier distribution 
is related to the couplings to low-lying collective vibrational states
and rotational states. The vibrational modes are connected to the change of 
nuclear shape. Nuclear rotational states are related to static deformations of 
the interacting nuclei. Furthermore, when the two nuclei come close enough to 
each other, both nuclei are distorted owing to the attractive nuclear force 
and 
the repulsive Coulomb force, thus dynamical deformations develop
\cite{Wang2012_PRC85-041601R,Zagrebaev2003_PRC67-061601R}.
Considering the dynamical deformation, a two-dimensional potential energy 
surface (PES) with respect to relative distance $R$ and quadrupole deformation 
of the system can be obtained. Based on the PES, empirical formulas
were proposed for calculating the parameters of the barrier distribution in 
Ref. 
\cite{Wang2015_arXiv1504.00756}. Note that such empirical formulas are 
connected with the quadrupole deformation parameters predicted by the 
finite-range droplet model (FRDM) \cite{Moller1995_ADNDT59-185}.

The effect of the coupling to the PQNT channel is simulated by broadening the 
barrier distribution. Only
one neutron pair transfer channel is considered in the ECC model. When the
$Q$ value for one neutron pair transfer is positive, the widths of the barrier
distribution are calculated as  $ \varDelta_i  \rightarrow  
fQ(2n)+\varDelta_i, (i=1,2)$, where $Q(2n)$ is the 
$Q$ value for one neutron pair 
transfer. $f$ is taken as $0.32$ for all reactions 
with positive $Q$ value for one neutron pair transfer channel 
\cite{Wang2015_arXiv1504.00756}.
 
\section{Results and Discussions} \vspace*{-1mm} 
\label{sec:results}

The fusion excitation function is influenced by two types of features related to
the structure of and interaction potential between the projectile and the target. 
One is of a static nature, such as the heights, curvatures and radii of the barriers, 
and the static effects associated with the excess protons or neutrons in weakly 
bound nuclei. The other one is the dynamic effect of couplings to inelastic 
excitations, the breakup channel, and nucleon transfer channels. In order to 
study the dynamic coupling effects on fusion
cross sections, it is necessary to eliminate the geometrical factors
and static effects of the potential between the two 
nuclei \cite{Canto2009_JPG36-015109,Canto2009_NPA821-51}. In the present 
work, we will first investigate 
the dynamic coupling effects on fusion cross sections by eliminating the 
static effects with the 
UFF formalism. Then, the effects of couplings to 
inelastic excitations and neutron transfer channels on fusion are analysed 
separately with the ECC model. Since $^{32}$S and $^{40}$Ca are both well bound, 
the breakup effects is not important.

\begin{figure}[H]
\centering{\includegraphics[width=0.49\columnwidth]{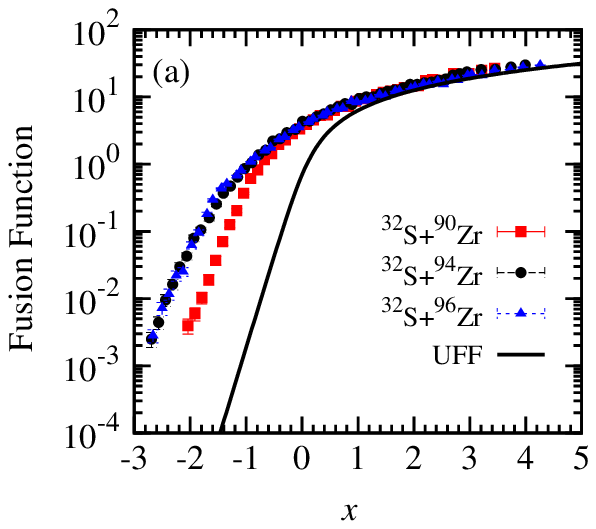}            
\includegraphics[width=0.49\columnwidth]{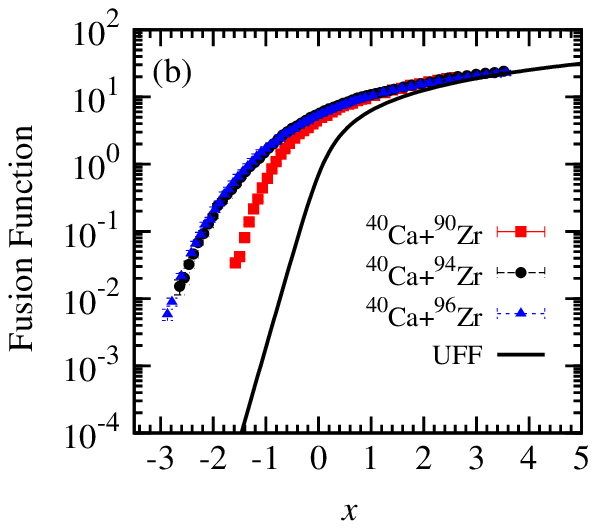}}
\caption{(Color online) The reduced fusion function for reactions 
(a) ${}^{32}$S + ${}^{90,94,96}$Zr and (b) 
${}^{40}$Ca + ${}^{90,94,96}$Zr as a function of $x$.
The solid line represents the UFF. The experimental fusion cross sections
are taken from
Refs.~\cite{Zhang2010_PRC82-054609,Timmers1998_NPA633-421,
Stefanini2007_PRC76-014610,Jia2014_PRC89-064605}.}\label{fig:red}
\end{figure}
\vspace*{-5mm}

\begin{figure*}[hbt!]
\centering{
\includegraphics[width=0.5\columnwidth]{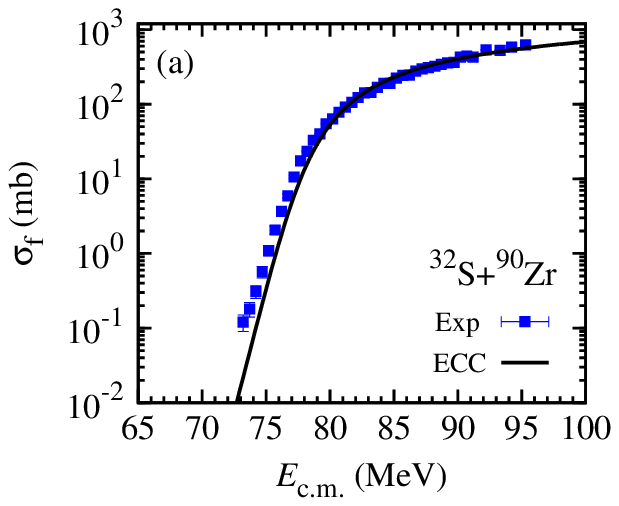}
\includegraphics[width=0.5\columnwidth]{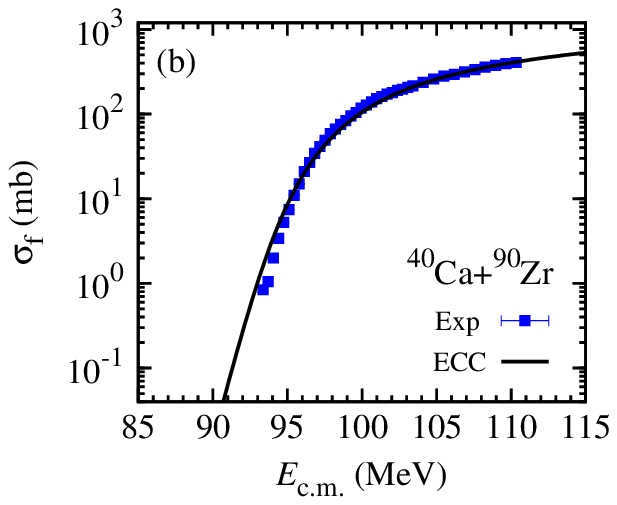} 
\includegraphics[width=0.5\columnwidth]{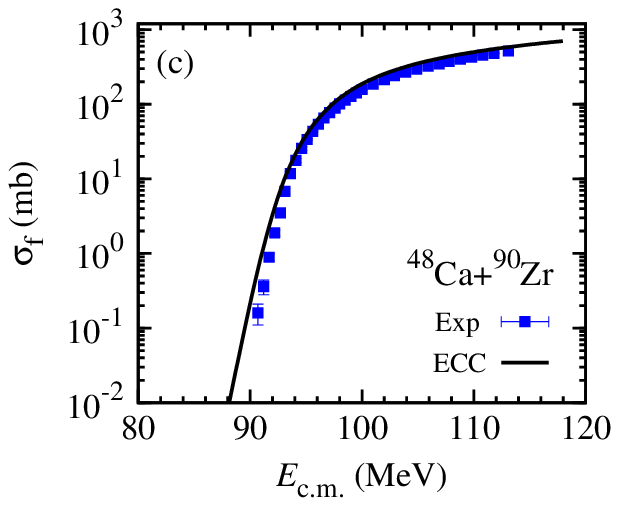}
\includegraphics[width=0.5\columnwidth]{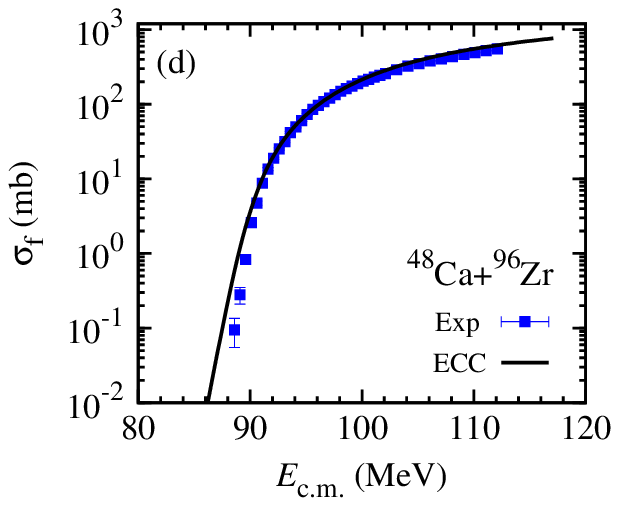}}
\caption{(Color online) The calculated and experimental fusion excitation 
functions for the reactions (a) ${}^{32}$S+ ${}^{90}$Zr, (b) ${}^{40}$Ca + 
${}^{90}$Zr, (c) ${}^{48}$Ca + ${}^{90}$Zr, and (d) ${}^{48}$Ca + ${}^{96}$Zr. 
The solid lines denote the 
results from the ECC calculations. The experimental fusion excitation 
functions are taken from
Refs.~\cite{Zhang2010_PRC82-054609,Timmers1998_NPA633-421,
Stefanini2006_PRC73-034606}. }\label{fig:NNT}
\end{figure*} 
 
\begin{figure*}[hbt!]
\centering{
\includegraphics[width=0.5\columnwidth]{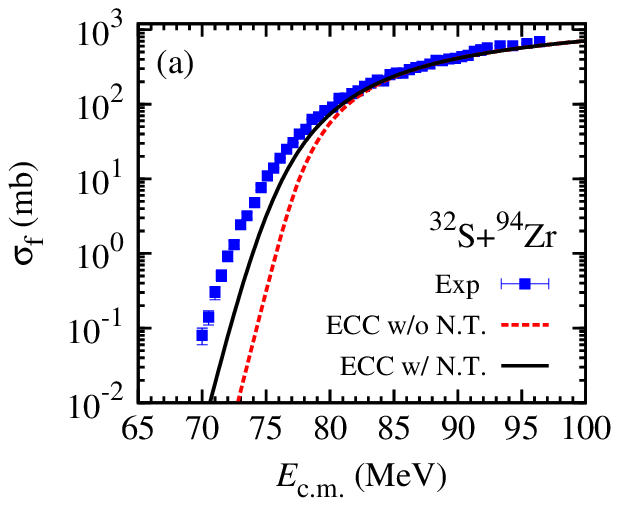}
\includegraphics[width=0.5\columnwidth]{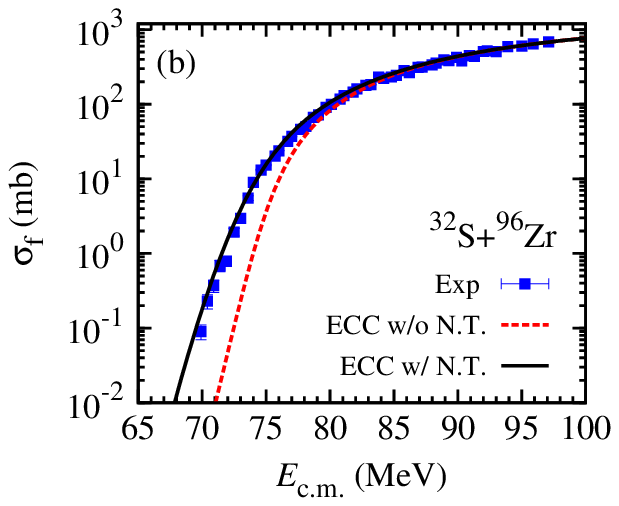} 
\includegraphics[width=0.5\columnwidth]{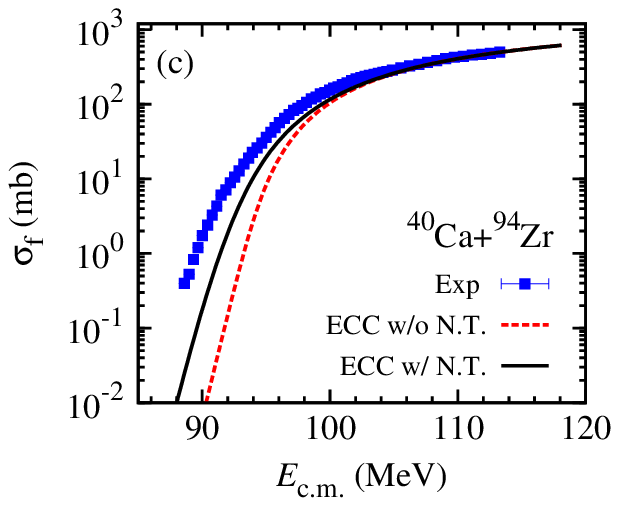}
\includegraphics[width=0.5\columnwidth]{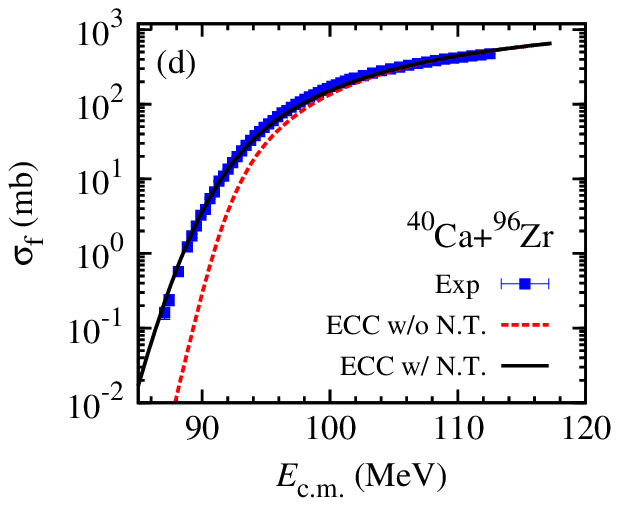}}
\caption{(Color online) The calculated and experimental fusion excitation 
functions for reactions (a) ${}^{32}$S + ${}^{94}$Zr,
(b) ${}^{32}$S + ${}^{96}$Zr, (c) ${}^{40}$Ca + ${}^{94}$Zr,
and (d) ${}^{40}$Ca + ${}^{96}$Zr. 
The dash lines denote the results from the ECC calculations 
without the coupling to the neutron transfer
channels considered. The 
solid lines denote the results from the ECC calculations with 
the coupling to the neutron transfer
channels considered. The quadrupole deformation 
parameters of $\beta_2=0.062$ and $\beta_2=0.217$ are used for $^{94}$Zr and 
$^{96}$Zr \cite{Moller1995_ADNDT59-185}, respectively. The experimental 
fusion excitation functions are taken from 
Refs.~\cite{Zhang2010_PRC82-054609,Jia2014_PRC89-064605,
Stefanini2007_PRC76-014610, Timmers1998_NPA633-421}.
}\label{fig:Zr9496}
\end{figure*} 

\subsection{\label{sec:ff}Reduced fusion excitation functions} 

We adopt the method
proposed in Refs.~\cite{Canto2009_JPG36-015109,Canto2009_NPA821-51} to 
eliminate the geometrical factors and static effects of the 
potential between the two nuclei. According
to this prescription, the fusion cross section and the collision energy are
reduced to a dimensionless fusion function $F(x)$ and a dimensionless variable
$x$,
\begin{equation}\label{eq:1}
F(x) = \frac{2E_{\rm c.m.}}{R_{\rm B}^2\hbar\omega}\sigma_{\rm F}, \quad
   x = \frac{ E_{\rm c.m.}-V_{\rm B}}{\hbar\omega},
\end{equation}
where $E_{\rm c.m.}$ is the collision energy
in the center-of-mass frame, $\sigma_{\rm F}$ is the fusion cross section,
and $V_{\rm B}$, $\hbar\omega$, and $R_{\rm B}$
denote the height, curvature, and radius of the barrier which is
approximated by a parabola. The barrier parameters $V_{\rm B}$, 
$\hbar\omega$, and $R_{\rm B}$ are obtained from the double folding and
parameter-free S\~ao Paulo potential (SPP)
\cite{CandidoRibeiro1997_PRL78-3270,Chamon1997_PRL79-5218,
Chamon2002_PRC66-014610}.

\vspace*{-5mm} 
\begin{table}[H]
\caption{The barrier heights, radii, and curvatures used to 
reduce the fusion excitation functions.}\label{tab:par_po}
\centering{\footnotesize
\doublerulesep 0.2pt 
\begin{tabular*}{0.4\paperwidth}{@{\extracolsep{\fill}}lccc}
\toprule
  Reaction & $V_{\rm B}$ & $\hbar\omega$ & $R_{\rm B}$   \\
   & (MeV) & (MeV) & (fm)   \\
 \hline 
 $^{32}$S$+^{90}$Zr  & 81.37  & 4.00  & 10.54  \\
 $^{32}$S$+^{94}$Zr  & 80.64  & 3.95  & 10.64  \\
 $^{32}$S$+^{96}$Zr  & 80.30  & 3.90  & 10.70  \\
 $^{40}$Ca$+^{90}$Zr & 99.94  & 4.02  & 10.74   \\
 $^{40}$Ca$+^{94}$Zr & 99.07  & 3.97  & 10.84   \\
 $^{40}$Ca$+^{96}$Zr & 98.65  & 3.97  & 10.88   \\
\bottomrule
 \end{tabular*} }
\end{table}
\vspace*{-3mm}

If the fusion cross section can be
accurately described by the Wong's formula \cite{Wong1973_PRL31-766}
\begin{equation}\label{eq:2}
\sigma_{\rm F}^{\rm W}(E_{\rm c.m.}) = \frac{R_{\rm B}^2\hbar\omega}
      {2E_{\rm c.m.}}\ln\left[1+\exp\left(\frac{2\pi(E_{\rm c.m.}-V_{\rm B})} 
      {\hbar\omega}\right)\right],
\end{equation}
then $F(x)$ reduces to
\begin{equation}\label{eq:3}
   F_{0}(x) = \ln\left[1+\exp(2\pi x)\right],
\end{equation}
which is called the universal fusion function (UFF)
\cite{Canto2009_JPG36-015109,Canto2009_NPA821-51}. Note
that $F_{0}(x)$ is independent of reaction systems. 
So $F_{0}(x)$ is used as a uniform standard reference to explore the coupling 
effects on fusion cross sections. Deviations of the fusion function from the 
UFF, if exist, are assumed to mainly arise from the dynamic coupling effects on 
fusion cross section.

The reduced fusion excitation functions of the reactions $^{32}$S + 
$^{90,94,96}$Zr and $^{40}$Ca + $^{90, 94,96}$Zr are shown
in Fig.~\ref{fig:red}. The solid lines represent the UFF. 
The parameters of the potential used in the reduction procedure 
are obtained from the SPP and listed in Table~\ref{tab:par_po}. 
On one hand, the deviations of the reduced fusion
excitation functions from the UFF for $^{32}$S + $^{90,94,96}$Zr 
and $^{40}$Ca + $^{90, 94,96}$Zr are very large, especially for  
$^{32}$S + $^{94,96}$Zr and $^{40}$Ca + $^{94,96}$Zr. This 
implies that the enhancement of the sub-barrier cross sections due to the 
coupling effects in $^{32}$S, $^{40}$Ca + $^{94,96}$Zr are much larger than 
that in $^{32}$S,$^{40}$Ca + $^{90}$Zr. This is because that the neutron 
transfer channels are opened in $^{32}$S, $^{40}$Ca + $^{94,96}$Zr. The $Q(xn)$ 
values for the PQNT channels from ground state to ground state for 
$^{32}$S + $^{90,94,96}$Zr and $^{40}$Ca + $^{90, 94,96}$Zr are shown in 
Table~\ref{tab:Qval}. On the other hand, the behaviors of the 
reduced fusion excitation functions of $^{32}$S + $^{94}$Zr and $^{32}$S + 
$^{96}$Zr are very similar. The situation is the same for the reactions 
$^{40}$Ca 
+ $^{94,96}$Zr. This implies that the total effect of the couplings to  
inelastic excitations and neutron transfer channels in $^{32}$S + 
$^{94}$Zr ($^{40}$Ca + $^{94}$Zr) is almost the same as that in $^{32}$S + 
$^{96}$Zr 
($^{40}$Ca + $^{96}$Zr).

It is well-known that the coupling to PQNT channels enhances the 
sub-barrier fusion cross section. However, a quantitative 
understanding of the enhancement due to PQNT channel coupling remains elusive
because the only observable is the total enhancement of the cross
section. In order to further understand the effect of the 
neutron 
transfer channels in these reactions, we 
will study the effects of couplings to inelastic excitations and neutron 
transfer channels on fusion separately by using the ECC model.

\subsection{Couplings to inelastic excitations and PQNT channels}\vspace*{-1mm}

In this section, we will isolate the effect of transfer coupling from 
that of couplings to the inelastic excitations 
channels. We first estimate the enhancement due to inelastic excitations channel 
couplings alone by adopting the ECC model \cite{Wang2015_arXiv1504.00756}.
 
The experimental fusion excitation functions and the results from ECC 
calculations for ${}^{32}$S,${}^{40,48}$Ca + ${}^{90}$Zr and 
${}^{48}$Ca + ${}^{96}$Zr are shown in Fig.~\ref{fig:NNT}. For these four 
reactions, the $Q$ values of 
neutron transfer are negative as seen in Table~\ref{tab:Qval}. Therefore, 
only the couplings to inelastic excitations channels are responsible for the 
enhancement. The solid lines 
denote the results from the ECC calculations. In the ECC calculations, the 
static quadrupole deformation parameters $\beta_2=0.035$, $\beta_2=0.062$, 
and $\beta_2=0.217$ are used for $^{90,94,96}$Zr \cite{Moller1995_ADNDT59-185}, 
respectively. One can find that results from the ECC   
calculations are in good agreement with the data.   
This implies that the ECC model with the barrier distributions obtained from 
Ref.~\cite{Wang2015_arXiv1504.00756} can describe well the effect of the 
couplings to inelastic excitations channels. Therefore, the ECC 
calculations can provide an accurate quantitative estimate
of the enhancement due to inelastic excitations channel couplings alone.

As mentioned above, the sub-barrier cross section of 
reactions $^{32}$S + $^{94,96}$Zr ($^{40}$Ca + $^{94,96}$Zr) shows an extra 
enhancement as compared with that of $^{32}$S + $^{90}$Zr ($^{40}$Ca + 
$^{90}$Zr). This is because that the PQNT channels are opened in 
${}^{32}$S ($^{40}$Ca) + ${}^{94,96}$Zr, c.f. the $Q$ values for neutron 
transfer listed in Table~\ref{tab:Qval}. We first perform
the ECC calculations with couplings to the inelastic 
excitations channels considered only. The comparison of the results from 
ECC calculations to the experimental values for ${}^{32}$S + ${}^{94,96}$Zr and 
${}^{40}$Ca + ${}^{94,96}$Zr are shown in Fig.~\ref{fig:Zr9496} by the dash 
lines. 
It can be seen that the experimental fusion data at near-barrier and 
sub-barrier energy show large enhancement as compared with the ECC calculations 
without the coupling to neutron transfer channels considered. 
Consequently, these enhancements may be from the coupling to PQNT channels. 
 
\begin{figure}[H]
\centering
\includegraphics[width=0.49\columnwidth]{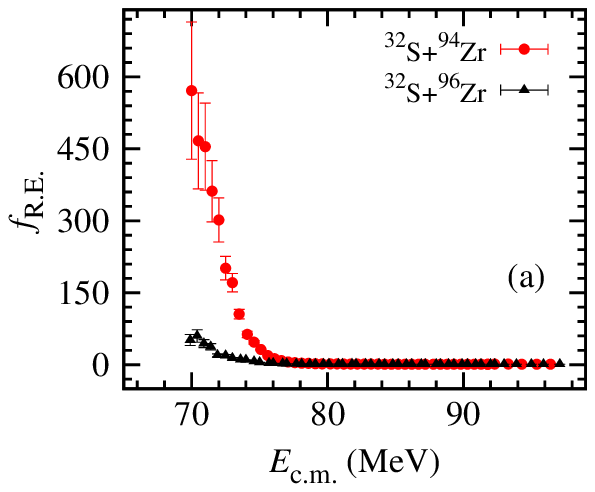}
\includegraphics[width=0.49\columnwidth]{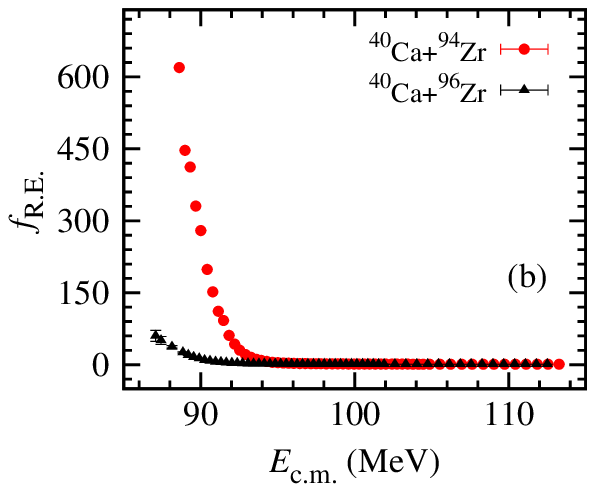}\\
\includegraphics[width=0.49\columnwidth]{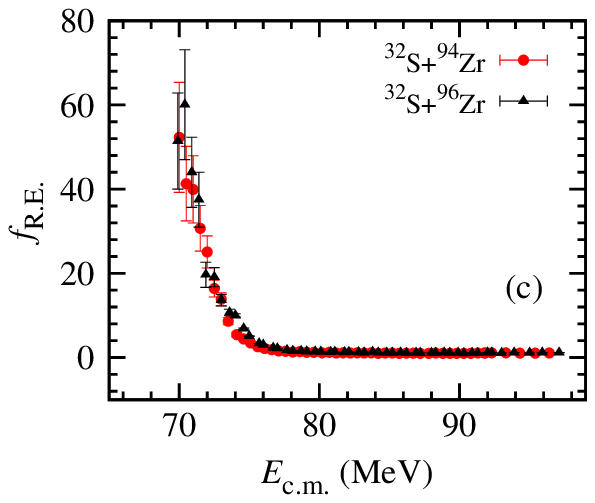}
\includegraphics[width=0.49\columnwidth]{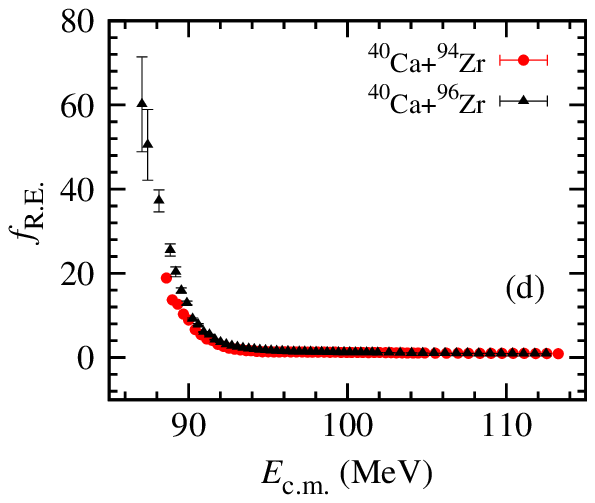}
\caption{(Color online) The relative enhancement $ f_{\rm R.E.}$ owing to the 
coupling to PQNT channels for the reactions ${}^{32}$S + ${}^{94,96}$Zr [(a) and (c)] and 
${}^{40}$Ca + ${}^{94,96}$Zr [(b) and (d)]. 
The upper panels (a) and (b) show the results obtained with $\beta_2=0.062$ for $^{94}$Zr 
used and the lower panels (c) and (d) show the results obtained with $\beta_2=0.217$ for 
$^{94}$Zr used.
The 
experimental fusion excitation functions are taken from 
Refs.~\cite{Zhang2010_PRC82-054609,Jia2014_PRC89-064605,
Stefanini2007_PRC76-014610,
Timmers1998_NPA633-421}.}\label{fig:R}
\end{figure}  
 
In the ECC model, the effect of coupling to the PQNT channels is simulated 
by broadening the barrier distribution which is related to the $Q(2n)$ value.  
The results from ECC calculations with the neutron transfer effects taken into 
account are shown in Fig.~\ref{fig:Zr9496} by the solid line.  
For $^{32}$S + $^{96}$Zr and $^{40}$Ca + $^{96}$Zr, it can be seen that 
the calculated results are in good agreement with the experimental values. 
But, for $^{32}$S + $^{94}$Zr and $^{40}$Ca + $^{94}$Zr, the calculated results 
underestimate the sub-barrier cross sections considerably. 
To understand this underestimation, we follow Jia {\it et al.} \cite{Jia2014_PRC89-064605}
and examine the relative enhancement. 
The relative enhancement is calculated as the ratio of the experimental fusion 
cross section to the calculated result by using the ECC model 
without the coupling to the neutron transfer channels considered, i.e., 
$ f_{\rm R.E.} = \sigma^{\rm exp}(E_{\rm c.m.}) / 
                 \sigma^{\rm th}_{\rm ECC}(E_{\rm c.m.})$. 
Figure~\ref{fig:R} shows these relative enhancements for 
${}^{32}$S + ${}^{94,96}$Zr and ${}^{40}$Ca + ${}^{94,96}$Zr. 
From Fig.~\ref{fig:R}(a) and Fig.~\ref{fig:R}(b), one can find that the relative 
enhancements for the reactions with $^{94}$Zr as target are much larger than 
those for the reactions with $^{96}$Zr as target. This implies that if the 
estimates of the enhancement due to inelastic excitations channel couplings for 
the ${}^{32}$S + ${}^{94}$Zr and ${}^{40}$Ca + ${}^{94}$Zr are reliable, 
the effect of PQNT in $^{32}$S + $^{94}$Zr (${}^{40}$Ca + ${}^{94}$Zr) is 
much stronger than that in $^{32}$S + $^{96}$Zr (${}^{40}$Ca + ${}^{96}$Zr).

\begin{figure}[H]
\centering{
\includegraphics[width=0.5\columnwidth]{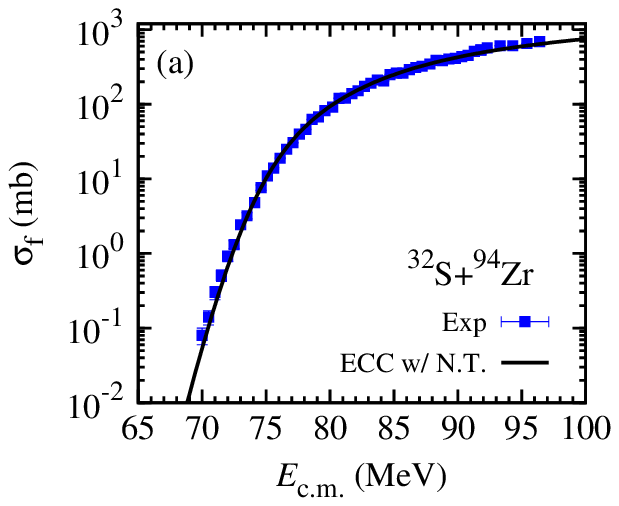}\hspace{-0.5em}
\includegraphics[width=0.5\columnwidth]{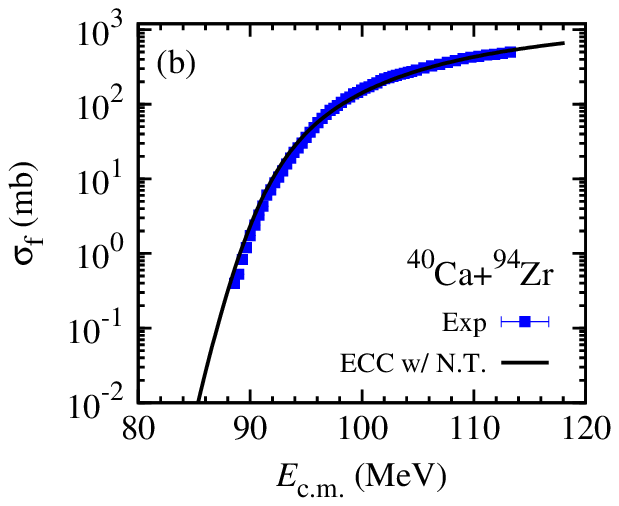}}
\caption{(Color online) The calculated and experimental fusion excitation 
functions for the reactions (a) ${}^{32}$S + ${}^{94}$Zr and (b) 
${}^{40}$Ca + ${}^{94}$Zr. The calculated 
fusion excitation functions are obtained from ECC 
calculations with the neutron transfer effects taken into 
account. The quadrupole deformation parameter $\beta_2=0.217$ for $^{94}$Zr is used. 
The solid squares denote the experimental values taken from 
Refs.~\cite{Jia2014_PRC89-064605,Stefanini2007_PRC76-014610}.}\label{fig:beta}
\end{figure}

Next let's discuss the discrepancies between the calculated fusion cross sections
and the experimental values from another viewpoint: We first estimate the
effect of coupling to the PQNT channels, then constrain the coupling effects
due to inelastic excitation channels.
Within the ECC model, for ${}^{32}$S + ${}^{94,96}$Zr and ${}^{40}$Ca + ${}^{94,96}$Zr, 
the influence of neutron transfer should be almost the same 
because $Q(2n)$ values for these four reactions are very similar (see Table~\ref{tab:Qval}). 
As discussed in Sec.~\ref{sec:ff}, 
the total effect of the couplings to the inelastic excitations 
and neutron transfer channels in $^{32}$S + $^{94}$Zr ($^{40}$Ca + 
$^{94}$Zr) is also almost the same as that in $^{32}$S + $^{96}$Zr ($^{40}$Ca + 
$^{96}$Zr). 
Therefore the enhancement of the fusion cross section due to couplings to 
inelastic excitations channels in reactions with $^{94}$Zr and $^{96}$Zr 
should be similar to each other. 
This implies that the structure information related to fusion as described by 
the ECC model for $^{94}$Zr and $^{96}$Zr should be similar, i.e., the quadrupole 
deformation parameters $\beta_2$'s for  $^{94}$Zr and $^{96}$Zr should be similar 
to each other in ECC calculations. 
In order to check this conjecture, we calculate the 
fusion cross sections with $\beta_2=0.217$ for $^{94}$Zr, the same as that of $^{96}$Zr. 
The results obtained from ECC calculations with 
the PQNT effect taken into account are shown in Fig.~\ref{fig:beta}. 
One can find that the calculated fusion cross sections are 
in good agreement with the data. 
In addition, as can be seen in Fig.~\ref{fig:R}(c) and Fig.~\ref{fig:R}(d),
the relative enhancements of the reactions with $^{94}$Zr as target are almost 
the same as those of the reactions with $^{96}$Zr as target. 

\subsection{The issue of quadrupole deformation for $^{94}$Zr}

Within the ECC model~\cite{Wang2015_arXiv1504.00756}, the formulas for 
calculating the parameters of the barrier distribution were proposed based on 
the quadrupole deformation parameters $\beta_2$ predicted by the FRDM 
\cite{Moller1995_ADNDT59-185}. 
According to the FRDM, $\beta_2=0.062$ for $^{94}$Zr and $\beta_2=0.217$ for $^{96}$Zr; 
they are quite different. 

As discussed before, the total effect of couplings to inelastic 
excitations and neutron transfer channels on fusion in the reaction 
$^{32}$S + $^{94}$Zr ($^{40}$Ca + $^{94}$Zr) is almost the same as that in 
the reaction $^{32}$S + $^{96}$Zr ($^{40}$Ca + $^{96}$Zr). 
On one hand, according to the estimate of the enhancement due to inelastic 
excitations channel couplings obtained from ECC calculations
with $\beta_2=0.062$ for $^{94}$Zr and $\beta_2=0.217$ for $^{96}$Zr, 
one may conclude that the role of PQNT channels in the reactions with $^{94}$Zr as 
target should be very different from that in the reactions with $^{96}$Zr as target. 
On the other hand, within the ECC model, the role of PQNT channels 
in ${}^{32}$S + ${}^{94,96}$Zr and ${}^{40}$Ca + ${}^{94,96}$Zr
should be similar because the $Q(2n)$ values are similar in these four reactions. 
This implies that the quadrupole deformation parameters of $^{94}$Zr and 
$^{96}$Zr should be similar to each other. 
Indeed, if one assumes that the quadrupole deformation parameter of $^{94}$Zr
is the same as that of $^{96}$Zr, i.e., $\beta_2=0.217$,
the fusion cross sections for reactions with $^{94}$Zr as target from the ECC 
model are in good agreement with the experiment (see Fig.~\ref{fig:beta}). 

Therefore, it becomes a puzzling issue whether the quadrupole deformation parameters 
for $^{94}$Zr and $^{96}$Zr are similar to each other or not.
In Table~\ref{tab:beta} we present $\beta_2$ values of $^{94}$Zr and $^{96}$Zr
given in Refs.~\cite{Raman2001_ADNDT78-1, Bhuyan2015_PRC92-034323}.
The quadrupole deformation parameters deduced from $B(E2; 
\textrm{g.s.}\rightarrow 2^+_1)$
for $^{94}$Zr and $^{96}$Zr are quite small, but very close to each other, 
$\beta_2(^{94}{\textrm{Zr}}) = 0.09$ and $\beta_2(^{96}{\textrm{Zr}}) = 0.08$ 
\cite{Raman2001_ADNDT78-1}.
In Ref.~\cite{Bhuyan2015_PRC92-034323}, a 
relativistic mean-field model was adopted to study the structural evolution 
in transition nuclei including $^{94}$Zr and $^{96}$Zr. With the NL3 
interaction, the obtained $\beta_2$'s for $^{94}$Zr and 
$^{96}$Zr are $0.169$ and $0.243$, which are also similar. 
However, with the NL3* interaction, the obtained $\beta_2$'s for $^{94}$Zr and 
$^{96}$Zr are $0.002$ and $0.233$, which are quite different. 

\vspace*{-5mm} 
\begin{table}[H]
\caption{The quadrupole deformation parameters for $^{94}$Zr and $^{96}$Zr.}\label{tab:beta}
\centering{\footnotesize
\doublerulesep 0.2pt 
\begin{tabular*}{0.4\paperwidth}{@{\extracolsep{\fill}}ccccc}
\toprule
  $\beta_2$ & FRDM~\cite{Moller1995_ADNDT59-185} & 
RMF (NL3)~\cite{Bhuyan2015_PRC92-034323} &  
RMF (NL3*)~\cite{Bhuyan2015_PRC92-034323} & Expt.~\cite{Raman2001_ADNDT78-1}  
\\ \hline 
$^{94}$Zr  & 0.062  & 0.169  & 0.002 & 0.09 \\
$^{96}$Zr  & 0.217  & 0.243  & 0.233 & 0.08\\ 
\bottomrule
 \end{tabular*} }
\end{table}
\vspace*{-3mm} 

To solve this puzzling issue and get further understanding of the coupling to 
PQNT channels, we suggest to measure the fusion 
excitation function of the reactions $^{48}$Ca + $^{94}$Zr or $^{36}$S + 
$^{94}$Zr. In these two reactions, the PQNT channels are closed (see Table 
\ref{tab:Qval}). Therefore, these two reactions can be used to test the 
structure information connected to fusion of the target 
$^{94}$Zr. If the results obtained from ECC calculations together with the 
quadrupole deformation parameters predicted from FRDM are in good 
agreement with the measured fusion excitation functions, one can conclude that 
the influence of PQNT channel coupling on sub-barrier fusion cross section in 
the reaction $^{32}$S + $^{94}$Zr ($^{40}$Ca + $^{94}$Zr) is stronger than 
that in the reaction $^{32}$S + $^{96}$Zr ($^{40}$Ca + $^{96}$Zr). 
Otherwise, further study of the structure related to fusion of $^{94}$Zr are needed. 
In any case, we can get a better understanding of the coupling to PQNT channels.
 
\section{Summary} \vspace*{-1mm} 
\label{sec:summary}
In summary, we adopt the universal fusion function formalism and the ECC model 
to investigate the 
dynamic coupling effects on fusion cross sections for the reactions 
$^{32}$S + $^{94,96}$Zr and $^{40}$Ca + $^{94,96}$Zr. The reduced  
fusion excitation functions show that the total 
effect of inelastic excitations and neutron transfer channel couplings on 
fusion in $^{32}$S + $^{94}$Zr ($^{40}$Ca + $^{94}$Zr) is 
almost the same as that in  $^{32}$S + $^{96}$Zr ($^{40}$Ca + $^{96}$Zr). 
Within the ECC model, the enhancements due to inelastic excitations channel 
couplings and neutron transfer channel coupling are evaluated 
separately. The results show that influences of 
neutron transfer in the reactions with $^{94}$Zr as target should be almost the 
same as  that in the reactions with $^{96}$Zr as target. This implies that the 
quadrupole deformation parameters of  $^{94}$Zr and $^{96}$Zr should be 
similar to each other. However, 
the quadrupole deformation parameters predicted from FRDM used in the ECC model 
are quite different. Experiments on $^{48}$Ca + $^{94}$Zr or $^{36}$S + 
$^{94}$Zr are suggested to solve the puzzling issue of whether
the quadrupole deformation parameters for $^{94}$Zr and $^{96}$Zr are similar
to each other or not. 

\vspace*{-2mm} \Acknowledgements{\bahao This work has been partly supported by 
the National Key Basic Research Program of China (Grant No. 2013CB834400), 
the National Natural Science Foundation of China (Grants 
No. 11175252, 
No. 11120101005, 
No. 11275248,
No. 11475115,
and
No. 11525524), 
and
the Knowledge Innovation Project of the Chinese Academy of Sciences (Grant No. 
KJCX2-EW-N01).
The computational results presented in this work have been obtained on 
the High-performance Computing Cluster of SKLTP/ITP-CAS and 
the ScGrid of the Supercomputing Center, Computer Network Information Center of 
the Chinese Academy of Sciences.}


\end{multicols}

\end{document}